# Abundances of Suprathermal Heavy Ions in CIRs during the Minimum of Solar Cycle 23

R. Bučík[1], U. Mall[1], A. Korth[1], G. M. Mason[2]

[1]*Max-Planck-Institut für Sonnensystemforschung, Katlenburg-Lindau, Germany*

email: bucik@mps.mpg.de

[2]*Applied Physics Laboratory, Johns Hopkins University, Laurel, MD, USA*

**Abstract** In this paper we examine the elemental composition of the 0.1 – 1 MeV/nucleon interplanetary heavy ions from H to Fe in corotating interaction regions (CIRs) measured by the SIT (Suprathermal Ion Telescope) instrument. We use observations taken on board the STEREO spacecraft from January 2007 through December 2010, which included the unusually long solar minimum following solar cycle 23. During this period instruments on STEREO observed more than 50 CIR events making it possible to investigate CIR ion abundances during solar minimum conditions with unprecedented high statistics. The observations reveal annual variations of relative ion abundances in the CIRs during the 2007-2008 period as indicated by the He/H, He/O and Fe/O elemental ratios. We discuss possible causes of the variability in terms of the helium focusing cone passage and heliolatitude dependence. The year 2009 was very quiet in CIR event activity. In 2010 the elemental composition in CIRs were influenced by sporadic solar energetic particle (SEP) events. The 2010 He/H and He/O abundance ratios in CIRs show large event to event variations with values resembling the SEP-like composition. This finding points out that the suprathermal SEPs could be the source population for CIR acceleration.

# 1. Introduction

Energetic ions in corotating interaction regions (CIRs) at 1 AU have elemental abundances very close to the fast solar wind composition, except for the overabundance of $^4$He and Ne (Mason *et al.*, 2008a). The He and Ne are overabundant in CIRs by factors of approximately 4 and 3, respectively. The abundances reported by Mason *et al.* (2008a) are based on the observations from a complete solar cycle period. The solar minimum CIR abundances have been reported in the earlier surveys (Richardson *et al.*, 1993; Mason *et al.*, 1997), but they are based on smaller number of events.

Recent observations of heavy ions in CIRs suggest that solar energetic particles (SEPs) may provide a seed population for CIR acceleration (*e.g.,* Mason, Desai, and Li, 2012). Observations of the abundance ratios in mixed CIR/SEP periods (Malandraki *et al.*, 2007; Bučík *et al.*, 2009a) showed corotating abundances consistent with SEP composition. The overabundance of $^3$He has been reported by Mason *et al.* (2008a), suggesting that the remnant impulsive flare ions are accelerated in CIRs. Other arguments were provided by Sanderson *et al.* (1995) and Torsti, Anttila, and Sahla (1999) showing an increase of ion intensities and hardening of the energy spectra in CIRs after solar transient events. Richardson, Mazur, and Mason (1998) suggested that the increase in these corotating events may instead be related to the change in the CIR configuration. Note that hardening of the spectra in the late-phase of the CIR events has been discussed by Reames *et al.* (1997) in terms of strengthening of the corotating shocks. Kocharov *et al.* (2003) developed a model of the propagation and acceleration of the SEPs in corotating compression regions in the solar wind. This model shows that CIRs at ~1 AU can re-accelerate SEPs.

The singly ionized interstellar He and Ne pickup ions provide another population which is accelerated in CIRs. The interstellar pickup He$^+$ contributes about 25 % of the total He and Ne$^+$ about 5 % of the total Ne population in CIRs at 1 AU (Chotoo *et al.*, 2000; Möbius *et al.*, 2002). However, these contributions are too small to explain overabundance of He and Ne in CIRs. Other singly ionized energetic heavy ions (C, O, Mg, and Fe) were not observed in CIRs (Mazur,

Mason, and Mewaldt, 2002; Möbius *et al.*, 2002). The interstellar pickup ions are created by ionization of the interstellar atoms as they approach the Sun (Möbius *et al.*, 1985). The singly ionized pickup ions can also be produced by an inner source located close to the Sun. The inner source pickup ions originate from neutral atoms in interplanetary dust grains concentrated near the Sun and have been discussed as a potential candidate that may contribute to the energetic particles accelerated in CIRs (Gloeckler *et al.*, 2000). Observations of multiply charged C ions and a virtually insignificant amount of singly charged ions in many CIR events have ruled out the possibility that the inner source of pickup ions contribute to the seed population for CIRs (Möbius *et al.*, 2002).

Although many compositional features of the suprathermal heavy ions in CIRs are known the relative contribution from different sources is not well understood. In this paper we present first observations of the elemental abundances of the suprathermal heavy ions from the *Solar TErestrial RElations Observatory* (STEREO). We examine event-to-event variations of He/H, He/O, NeS/O and Fe/O abundance ratios in CIRs over the long solar minimum period between January 2007 and December 2010.

## 2. Observations

The two STEREO spacecraft (S/C) were launched into heliocentric orbit near the ecliptic plane in October 2006, STEREO-A preceding the Earth and STEREO-B trailing behind. The angular separation of the two STEREO S/C increases gradually in time and reached ~180° at the end of our survey period. The measurements presented here were made with the Suprathermal Ion Telescope (SIT) instrument (Mason *et al.*, 2008b) onboard the STEREO-A S/C. The SIT instrument is a time-of-flight mass spectrometer which measures ions from H to Fe in the energy range from 20 keV/n to several MeV/n.

Figure 1 provides an overview of the data over the period from January 2007 to December 2010. During the survey period the monthly sunspot number, produced by Solar Influences Data Analysis Center (SIDC) team, never exceed number 25, indicating a low solar activity level. The average value of the monthly sunspot numbers in the 2007-2009 period was only 5 and increased to 17 in 2010. Figure 1a shows suprathermal hourly averaged He ion intensity (particles/cm$^2$ s sr MeV/n) for 193, 386, and 773 keV/n measured by SIT-A. Figures 1b - 1e show event-integrated abundance ratios He/H, He/O, NeS/O and Fe/O. Shown are abundances of those events where 1 h averaged 193 keV/n He ion intensities exceed 5 particles/(cm$^2$ s sr MeV/n). We use SIT pulse-height data to determine the abundance ratios. The horizontal dashed lines present average abundances for CIR (Mason *et al.*, 1997, 2008a), large solar energetic particle (LSEP) (Mazur *et al.*, 1993; Desai *et al.*, 2006), impulsive SEP (ISEP) (Mason *et al.*, 2004) and interplanetary shock (IP) (Desai *et al.*, 2003) events listed in Table 1 (columns 4-9). The average NeS/O abundance denoted by dashed lines in Figure 1d was derived by summing the Ne, Mg, Si, and S relative abundances from the cited works. Typical abundances for ISEP and IP shock events are not shown for He/H ratio in Figure 1. The approximate value of He/H ratio in ISEP events of about ~ 0.1 has been provided by Reames (1993). The compositional variations of low energy ions in IP shock events have been studied for a long time (*e.g.*, Klecker *et al.*, 1981). Generally, the decrease of the He/H ratio near the passage of the shock was reported in these studies.

The filled circles in Figure 1 indicate CIR events. We identify CIR events using the list of the CIRs compiled by the STEREO magnetometer team at the University of California Los Angeles (http://www-ssc.igpp.ucla.edu/forms/stereo/stereo_level_3.html), based on plasma and magnetic field data. The CIR events in the period January 2007-September 2009 were discussed by Mason *et al.* (2009) and Bučík *et al.* (2009b, 2011). Columns 2 and 3 in Table 1 contain CIR unweighted average abundance ratios measured in this survey for 137 keV/n. The ratios were averaged separately for the period January 2007-December 2009 and for 2010. Owing to the instrument energy threshold (240 keV) the survey average He/H ratio was measured at 386 keV/n. In the figure, crosses mark events that were accompanied by a sharp rise in the intensity of solar relativistic electrons. Some of these events were associated with ICME (Interplanetary Coronal Mass Ejection) driven IP shocks. The list of the STEREO solar electron events has been compiled by Solar Electron and Proton Telescope (SEPT) instrument team at Universität Kiel (http://www2.physik.uni-

kiel.de/stereo), and the list of ICME driven shocks by the STEREO magnetometer team at the University of California Los Angeles (http://www-ssc.igpp.ucla.edu/forms/stereo/stereo_level_3.html). Considering the elemental abundances derived from the SIT measurement, the intensity increases marked by crosses are likely SEP events. We note that a few other weak SEP events occurred in the 2007-2010 period (Leske *et al*., 2010; Wiedenbeck *et al*., 2010) with the intensities below the selection threshold used here.

Table 2 lists SEP events observed on STEREO-A in the period 2007-2010. Columns 2-3 show the approximate start time of the SEP event as determined by the 193 keV/n hourly average He ion intensity. Column 4 lists the peak He intensity at 193 keV/n. Columns 5-8 contain event-integrated abundances. Column 9 has some notes on the SEP events. Column 9 notes SEP events which are listed in the automatically generated energetic proton event list from the STEREO LET (Low-Energy Telescope) instrument (http://www.srl.caltech.edu/STEREO/DATA/LET/Public/EventList/LETEventList.txt). Note that event number 5 is not in either the LET or SEPT SEP event lists. However, in this event the SIT observations showed ion velocity dispersion at energies below 1 MeV/n and the SEP-like composition. Event number 3 in Table 2 is marked as the SEP event in the LET and SEPT event lists. However, SIT observations showed that the event is characterized by the CIR-like elemental ratios (see Figure 1), suggesting that corotating intensities dominated at lower energies. In addition, event number 3 was accompanied by a CIR, included in the previously cited UCLA list of CIRs. This provides further support for the above characterization of this event in the SIT energy range.

Figure 1 shows that the CIR event He/H ratios during the 2007-2009 are consistent with the average He/H ratio observed in the previous solar minimum (Mason *et al*., 1997). Large event-to-event variations of He/H in the CIRs occurred in 2010 when SEP event activity increased considerably. In 2007-2009 the CIR event He/O ratios were close to the average He/O ratio obtained in the earlier surveys (Mason *et al*., 1997, 2008a). The CIR event He/O ratio showed also larger spread in 2010. In contrast to the He/H and He/O there is no noticeable increase in the scatter of the CIR Fe/O ratio in 2010. Note in Figure 1 that the CIR NeS/O ratio stays relatively constant in 2010 and shows less variation compared to the period 2007-2009. The ranges of Fe/O and He/O are a factor of ~ 8, and the range of NeS/O is a factor of ~ 4. The ranges of Fe/O and He/O well agree with the values reported by Mason *et al*. (2008a) while the range of NeS/O is about two times lower in this survey.

**2.1 CIR abundances in 2007-2009**

In Figure 2 we explore in more detail variations of the CIR event-integrated elemental ratios in the period January 2007-December 2009. Figure 2a shows the 137 keV/n Fe/O ratio; Figure 2b the 193 keV/n He/O ratio; Figure 2c the 386 keV/n He/H ratio. The first gray region denotes the period of approximately October 2007-February 2008 when the He/H ratio shows a local increase. During the most time of this period the Fe/O ratio was decreased. The narrow gray shaded area marks another period from October 2008 to November 2008 when He/H shows an increase. Because of the small number of events the features in the second period are less clear. Furthermore, the He/H ratio shows decrease from the beginning of 2007 to the middle of 2007. At the beginning of 2007 when He/H was enhanced the Fe/O was again depleted. The thick red horizontal lines indicate periods of the helium focusing cone traversals from Drews *et al.* (2010). Blue curves show 5-point running averages of the data. There are 7 events in the 1st focusing cone and 5 events in the 2nd cone (red lines) periods. We considered these numbers for selecting the width of the averaging window.

**2.2 CIR abundances in 2010**

In the following we examine how the increased solar activity with the appearing of the first larger SEP events influenced the composition in the energetic ions associated with the CIRs.

Figure 3 compares elemental abundances for several CIR events in 2008 and 2010. Panel 3a shows SIT-A 1 hour He ion intensity in five energy channels. Panel 3b shows solar wind speed from the PLASTIC instrument on STEREO-A (Galvin *et al*., 2008). Panels 3c – 3f show 1 hour averages of He/H, He/O, NeS/O and Fe/O abundance ratios. The gray shaded areas indicate the compression regions from the previously cited list of CIRs. The CIR events in February-March 2008 (left side in Figure 3) have typical corotating elemental abundances while the CIR events in March-April 2010 (right side) have He/H and He/O ratios decreasing to values typical of SEP abundances. The NeS/O and Fe/O ratios in March-April 2010 remained near corotating values and essentially do not differ from the abundances observed in February-March 2008.

The CIR events with SEP-like He/H ratio are shown in Table 3. Columns 2-3 in Table 3 list the approximate start time of the event. Columns 4-7 show the event-integrated abundances. The 386 keV/n He/H ratio for events in Table 3 is close to the average LSEP He/H ratio given by Mazur *et al*. (1993). The He/O ratio in the events listed in Table 3 is close to the average LSEP He/O ratio from the survey of Desai *et al.* (2006). We note that CIR event 5 in Table 3 (14 June 2010) occurred during the decay phase of the preceding small SEP event (12 June 2010) observed by the LET and the SEPT instruments. The SIT instrument at energies below 1 MeV/n did not measure any ion increase associated with this SEP event.

The SIT measurements show some differences between survey-averaged CIR abundances in the very quiet period 2007-2009 and in 2010 (see columns 2 and 3 in Table 1). In 2010 the He/H and He/O ratios are lower than 2007-09 by factors of ~ 0.5 and ~ 0.25, respectively, while Fe/O in 2010 is higher than 2007-09 a factor of 1.5. No significant difference is observed in NeS/O ratio. Richardson *et al*. (1993) reported that the abundances in CIR events indicate transition from SEP-like at solar maximum to so-called solar minimum corotating event (SMCE)-like at solar minimum. The authors also found that corotating events in slower solar wind streams have SEP-like and in higher speed streams SMCE-like abundances. As the slower streams were mostly present near solar maximum and higher speed streams at solar minimum they argued that the transition in CIR abundances reflects solar wind speed dependence and no solar activity. We note that the dependence of the CIR-event abundances on solar wind speed still remains unclear (Mason *et al*., 2008a). While Richardson *et al*. (1993) reported strong correlation with the solar wind speed for the He/O ratio and no correlation for the Ne/O ratio, Mason *et al*. (1997, 2008a) provided contradictory findings.

## 3. Discussion and conclusions

Using data from the PLASTIC instrument aboard STEREO-A, Drews *et al.* (2010) reported on enhancements of $He^+$ and $Ne^+$ pickup ions during the helium focusing cone traversal around 6 November 2007 and 1 October 2008 with an approximate half width of 54 days. The authors observed that the He focusing cone was more pronounced on 6 November 2007 than on the second passage around 1 October 2008. The first simultaneous observations of $He^+$ pickup ions in the focusing cone with the two STEREO S/C were presented by Möbius *et al.* (2010). The focusing cone is a region on the downwind side of the Sun where the interstellar neutral He is attracted by the Sun's gravitational field (Möbius *et al.*, 1985). In contrast to He, interstellar H does not form a focusing cone since the repulsion due to radiation pressure dominates the gravitational attraction.

The SIT-A observations showed that the periods of He/H ratio increase, marked by shaded areas in Figure 2, occurred near the focusing cone traversals. This suggests that the enhanced He/H in the CIRs may be due to pickup $He^+$ enhancing the He content of the CIR seed population. The pattern observed by SIT-A remained until January-February 2008, exceeding the period of the pickup ions enhancement observed by Drews *et al.* (2010). In addition to the He/H enhancements at the end of 2007 and at the end of 2008 the measurements showed other enhancements near the beginning of 2007. The timing of all three increases agrees well with the yearly passage of the He focusing cone (Gloeckler *et al.*, 2004).

We note here that the helium focusing cone shapes were completely different in the STEREO-A and STEREO-B crossings at the end of 2007 (Möbius *et al.*, 2010). Möbius *et al.* (2010) observed a distinct maximum of $He^+$ rate on STEREO-A and two maxima on the STEREO-B with a minimum at the expected location of the cone center. Due to the complicated cone structure the variations of the elemental ratios measured by STEREO-B would require a more detailed analysis.

Kallenbach *et al*. (2000) concluded that the suprathermal $He^+/He^{2+}$ abundance ratio associated with CIR events reflects the annual variations of the $He^+$ pick-up ion source population. In contrast, Klecker *et al*. (2001), Möbius *et al*. (2002) and Kucharek *et al*. (2003) have not found evidence of the gravitational focusing cone in observations of the $He^+/He^{2+}$ ratio in CIR energetic particles. The authors concluded that injection and acceleration conditions masked $He^+$ pickup ion variations. Note the observations reported in these studies were acquired over the period relatively close to the sunspot maximum of the solar cycle 23 and include also periods of ICMEs. In contrast, the observations reported in this survey were performed during prolonged solar minimum period dominated by stably recurring CIRs (Mason *et al*., 2009; Bučík *et al*., 2011). This could lead to the much more uniform injection and acceleration conditions in the CIRs making it possible to see the signature of the focusing cone.

The PLASTIC/STEREO-A investigations showed that $O^+$ pickup ions were distributed evenly in time and do not show any enhancement during focusing cone traversals (Drews *et al.*, 2010). Therefore, the variations of the Fe/O observed by the SIT-A are likely not related to $O^+$ pickup ions. This brings the question whether the enhancements in the He/H and He/O observed during the focusing cone passages were really due to contribution of the $He^+$ as both He/H and He/O correlate with the Fe/O ratio in the cone. Thus, other factors which may cause the variations in the CIR abundances, possibly on a yearly basis, should be investigated.

For example, the heliographic latitude of the STEREO S/C shows annual variations due to the tilt between the solar rotation axis and the normal of the ecliptic. Throughout 2007-2008, STEREO-A reached the maximum northern latitude in the middle of August 2007/2008 and maximum southern latitude near the end of February 2007/2008. At the beginning of November 2007/2008 and in the middle of May 2007/2008, STEREO-A crossed the heliographic equator. Incidentally, the November crossings occurred near the center of the helium focusing cone. It is interesting that time profile of the He/H in Figure 2c shows a local minimum approximately in August 2007, near the S/C maximum northern latitude, and local maximum approximately in February 2008, near the S/C maximum southern latitude. Zieger and Mursula (1998) reported on annual variations in the solar wind speed observed around the solar minima. The authors suggested that the variations are related to the Earth's orbital motion in a latitudinal asymmetric solar wind speed. Similarity in the variations of the solar wind speed and the CIR event abundances suggests that the annual variations observed by STEREO-A might also be interpreted in terms of the solar wind speed dependence.

We found a number of CIR events in 2010 with decreased He/H and He/O abundance ratios to the SEP composition while the NeS/O and Fe/O ratios remained close to the CIR abundances. The abundances in these CIRs show some similarity with the IP shock elemental composition. For three CIR events marked as SEP-like (events 3, 5, 7 in Table 3) the He/H ratio decreased below the typical LSEP values, which is characteristic for ICME-driven IP shock events. In addition, the Fe/O ratios, remaining near CIR-like values, might also resemble IP shock composition as both abundances are quite similar. The SEP events 6, 9, 10 and 13 (Table 2) were accompanied by the ICME-driven IP shocks. These events have the Fe/O ratio near the average CIR value (or value between the typical CIR and IP shock abundances) and the He/H below the average LSEP elemental ratio. The elemental composition of these IP shock events is remarkably similar to the composition of the above discussed CIR events.

The observations indicate that the changes in the CIR composition appeared in the period of the enhanced SEP activity, where CIRs may re-accelerate particles from preceding SEP events. For example, the CIR events number 5, 6, and 7, listed in Table 3, were preceded by the SEP events. The examination of the events in Table 2 and 3 shows that the 31 October 2010 CIR event (number 6) was preceded by the 24 October 2010 SEP event; the 19 December 2010 CIR event (number 7) was preceded by the 13 December 2010 SEP event. The observations by the SIT instrument show that both SEP events decayed near the beginning of the CIR event. As mentioned in Section 2.2, the 14 June 2010 CIR event (number 5) was also preceded by a small 12 June 2010 SEP event which was not observed at the SIT energies. Thus in these cases we are likely summing CIR particles and the contributions from SEPs would be negligible. The CIRs which were not immediately preceded by the SEP event could accelerate particles from so-called reservoirs of SEPs in the heliosphere. The outer boundaries of the reservoirs can be formed by CIRs (Lario *et al.*, 2003) or merging ICMEs (Roelof *et al.*, 1992) creating the structures with temporarily trapped SEPs whose intensities and frequencies were enhanced in 2010.

**Acknowledgements** This work was supported by the Bundesministerium für Wirtschaft under grant 50 OC 0904. The work at the Johns Hopkins University/Applied Physics Laboratory was supported by NASA under contract SA4889-26309 from the University of the California Berkeley and by National Science Foundation grant 0962653.

**Table 1.** Heavy ion abundances.

| | 2007-09 CIR[a] | 2010 CIR[b] | CIR[c] | CIR[d] | LSEP[e] | LSEP[f] | ISEP[g] | IP shock[h] |
|---|---|---|---|---|---|---|---|---|
| | (137 keV/n) | (137 keV/n) | (150 keV/n) | (385 keV/n) | (385 keV/n) | (>300 keV/n) | (385 keV/n) | (750 keV/n) |
| He/H | 0.153±0.006 | 0.080±0.014 | [0.125±0.065] | … | … | [0.032±0.003] | … | … |
| He/O | 291±9 | 219±43 | 113±20 | [273±72] | [75.0±23.6] | 52±4 | [54±14] | [44.4±14.4] |
| NeS/O | 0.356±0.017 | 0.387±0.016 | 0.48±0.13 | [0.477±0.017] | [0.675±0.020] | … | [1.158±0.022] | [0.678±0.014] |
| Fe/O | 0.063±0.004 | 0.096±0.013 | [0.08±0.03] | 0.088±0.007 | [0.404±0.047] | 0.24±0.03 | [0.95±0.005] | [0.236±0.01] |

Note: The ratios in square brackets correspond to the dashed lines in Figure 1.

[a] This work; average of 42 STEREO-A CIR events between January 2007 and December 2009. Fe/O is averaged over 40 CIR events. He/H is for 386 keV/n (see text).

[b] This work; average of 14 STEREO-A CIR events in 2010. He/H is for 386 keV/n (see text).

[c] Average of 17 CIR events during solar minimum between December 1992 and July 1995 (Mason *et al.*, 1997).

[d] Average of 41 CIR events during Solar cycle 23 between November 1, 1997 and June 1, 2007 (Mason *et al.*, 2008a).

[e] Average of 64 LSEP events between November 1997 and January 2005 (Desai *et al.*, 2006).

[f] Average of 10 LSEP events between late 1977 and early 1981 (Mazur *et al.*, 1993).

[g] Average of 20 ISEP events between September 1997 and April 2003 (Mason *et al.*, 2004).

[h] Average of 72 IP shocks between October 1997 and September 2002 (Desai *et al.*, 2003).

**Table 2.** SEP events observed on STEREO-A [a].

| SEP Event | Year | Start Time (Month Day/UT) | Peak Intensity (1/cm² s sr MeV/n) | He/H x 10 386 keV/n | He/O 193 keV/n | NeS/O 137 keV/n | Fe/O 137 keV/n | Notes |
|---|---|---|---|---|---|---|---|---|
| 1 | 2007 | May 23/17:30 | 6.16 ± 0.65 | 0.400 ± 0.045 | 85 ± 20 | 0.948 ± 0.411 | 0.762 ± 0.343 | b, c, d |
| 2 | 2009 | Nov 02/01:20 | 7.26 ± 0.70 | 3.683 ± 0.226 | 137 ± 30 | 0.306 ± 0.099 | 0.457 ± 0.128 | b, c |
| 3 | 2010 | Jan 17/02:00 | 23.4 ± 1.2 | 1.220 ± 0.035 | 180 ± 22 | 0.328 ± 0.071 | 0.117 ± 0.037 | b, c |
| 4 | 2010 | Feb 12/17:00 | 56.2 ± 1.9 | 0.409 ± 0.007 | 72 ± 3 | 0.679 ± 0.047 | 0.291 ± 0.026 | b, c, e |
| 5 | 2010 | Apr 16/00:00 | 14.9 ± 1.0 | 0.210 ± 0.012 | 65 ± 6 | 0.341 ± 0.047 | 0.197 ± 0.033 | |
| 6 | 2010 | Aug 18/12:00 | 2155 ± 50 | 0.180 ± 0.001 | 79 ± 1 | 0.521 ± 0.009 | 0.140 ± 0.004 | b, c, e, f |
| 7 | 2010 | Aug 24/00:00 | 9.39 ± 0.79 | 1.450 ± 0.041 | 48 ± 3 | 1.008 ± 0.109 | 1.060 ± 0.113 | b |
| 8 | 2010 | Aug 30/14:50 | 27.5 ± 1.3 | 1.613 ± 0.010 | 124 ± 5 | 0.851 ± 0.056 | 0.602 ± 0.042 | b, c |
| 9 | 2010 | Sep 06/16:00 | 123 ± 3 | 0.116 ± 0.001 | 90 ± 2 | 0.411 ± 0.018 | 0.070 ± 0.006 | b, c, e |
| 10 | 2010 | Sep 10/06:00 | 1486 ± 28 | 0.228 ± 0.002 | 90 ± 1 | 0.488 ± 0.011 | 0.138 ± 0.005 | b, c, e |
| 11 | 2010 | Oct 24/09:00 | 17.2 ± 1.1 | 1.288 ± 0.036 | 92 ± 8 | 0.878 ± 0.125 | 0.754 ± 0.110 | b |
| 12 | 2010 | Nov 26/20:00 | 5.00 ± 0.59 | 0.366 ± 0.017 | 52 ± 7 | 0.893 ± 0.235 | 0.249 ± 0.093 | b, c |
| 13 | 2010 | Dec 13/21:30 | 21.7 ± 1.2 | 0.110 ± 0.004 | 89 ± 8 | 0.370 ± 0.048 | 0.072 ± 0.017 | b, c, e |

[a] The peak intensities (1 h averaged) are for 193 keV/n He.

b – in STEREO SEPT solar electron event list

c – in STEREO LET SEP event list

d – event discussed by Chollet *et al.* (2010)

e – ICME IP shock; f – event discussed by Leske *et al.* (2011) and Gómez-Herrero *et al.* (2011)

**Table 3.** CIR events with SEP-like He/H ratios

| CIR Event | Year | Start Time (Month Day/UT) | He/H x 10 386 keV/n | He/O 193 keV/n | NeS/O 137 keV/n | Fe/O 137 keV/n |
|---|---|---|---|---|---|---|
| 1 | 2010 | Mar 12/03:00 | $0.585 \pm 0.044$ | $350 \pm 73$ | $0.547 \pm 0.166$ | $0.052 \pm 0.038$ |
| 2 | 2010 | Mar 15/03:00 | $0.446 \pm 0.007$ | $105 \pm 4$ | $0.350 \pm 0.024$ | $0.051 \pm 0.008$ |
| 3 | 2010 | Mar 21/05:00 | $0.139 \pm 0.003$ | $68 \pm 4$ | $0.375 \pm 0.033$ | $0.120 \pm 0.016$ |
| 4 | 2010 | Apr 08/03:00 | $0.349 \pm 0.012$ | $63 \pm 6$ | $0.363 \pm 0.059$ | $0.042 \pm 0.016$ |
| 5 | 2010 | Jun 14/11:00 | $0.248 \pm 0.008$ | $52 \pm 4$ | $0.334 \pm 0.052$ | $0.099 \pm 0.024$ |
| 6 | 2010 | Oct 31/05:30 | $0.368 \pm 0.015$ | $107 \pm 11$ | $0.363 \pm 0.057$ | $0.115 \pm 0.027$ |
| 7 | 2010 | Dec 16/08:00 | $0.167 \pm 0.008$ | $120 \pm 13$ | $0.421 \pm 0.066$ | $0.109 \pm 0.028$ |

**Figure 1.** (a) SIT/STEREO-A 1 h averaged He intensity (particles/cm$^2$ s sr MeV/n) for 193, 386, and 773 keV/n. (b - e) Event-averaged 386 keV/n He/H, 193 keV/n He/O, 137 keV/n NeS/O and 137 keV/n Fe/O elemental ratios. CIR events are marked by *filled circles*, SEP events by *crosses*. *Dashed lines* show abundances measured in various particle populations present in the heliosphere: CIR events (*blue*), LSEP events (*green*), ISEP events (*yellow*) and IP shocks events (*red*). The NeS/O ratio in LSEP and IP shock events are very similar and therefore the corresponding horizontal lines overlap each other.

**Figure 2.** CIR event-integrated abundance ratios in the period 2007-2009. (a) Fe/O for 137 keV/n. (b) He/O for 193 keV/n. (c) He/H for 386 keV/n. *Blue curves* indicate 5-point running averages of the data. *Shaded areas* and *red thick lines* are described in the text.

**Figure 3.** Panel (a) SIT/STEREO-A 1 h averaged He intensity (particles/cm$^2$ s sr MeV/n) for 193, 273, 386, 546 and 773 keV/n. Panel (b) Solar wind speed. Panels (c - f) 1 h averaged 386 keV/n He/H, 193 keV/n He/O, 137 keV/n NeS/O and 137 keV/n Fe/O elemental ratios. *Grey shaded regions* mark the time intervals of the CIRs. *Dashed lines* present abundances in various particle populations.

Figure 1

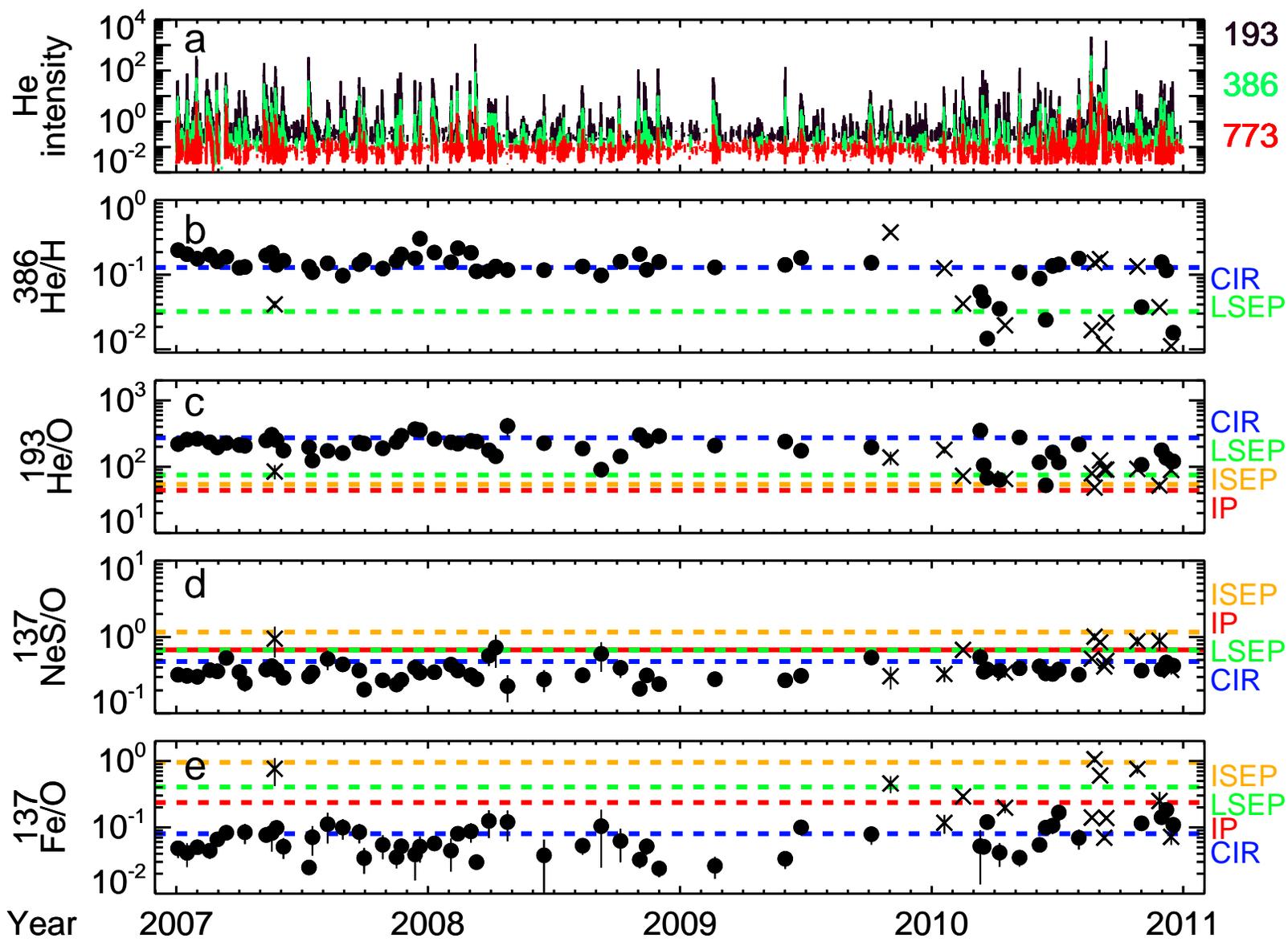

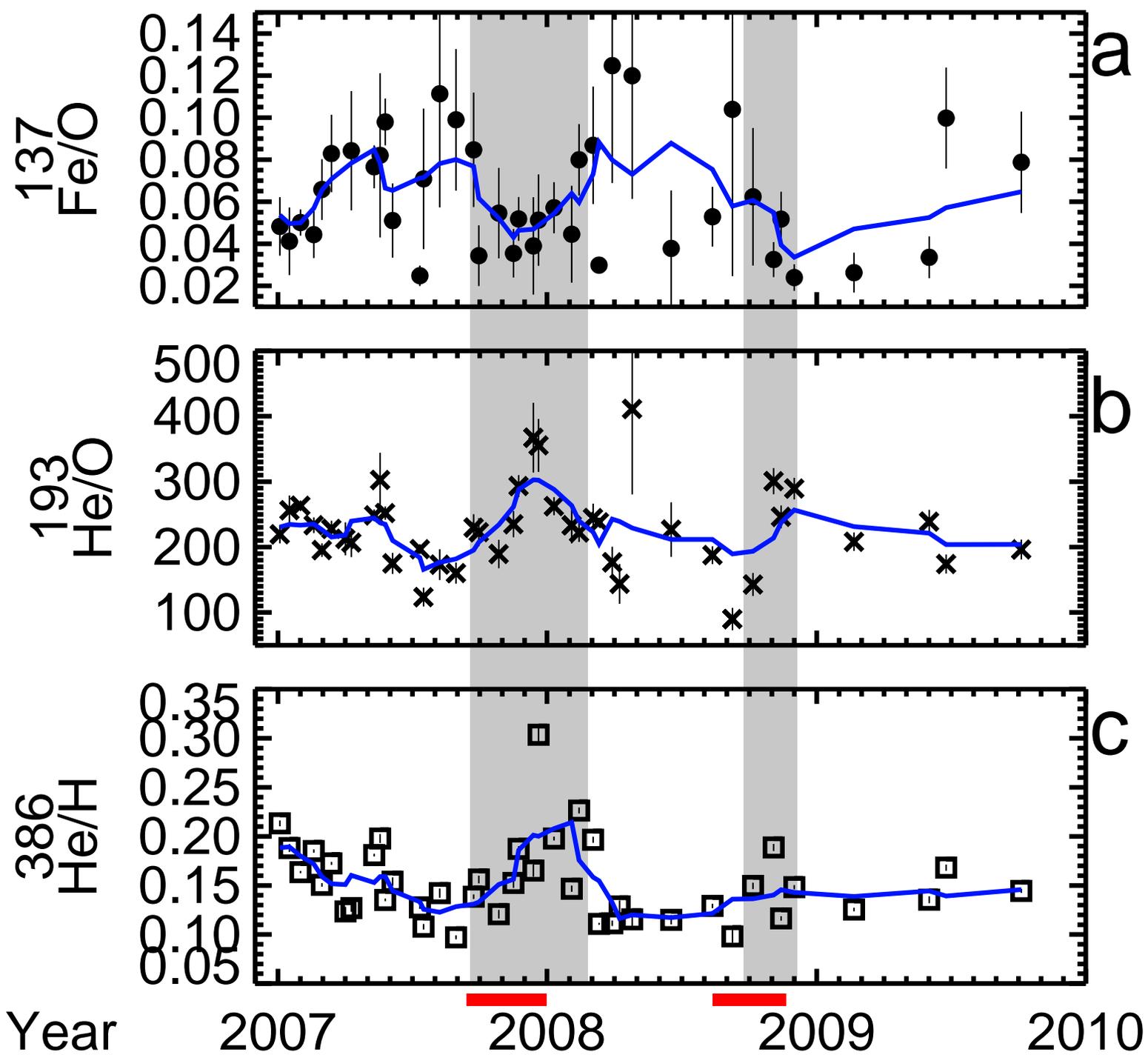

Figure 2

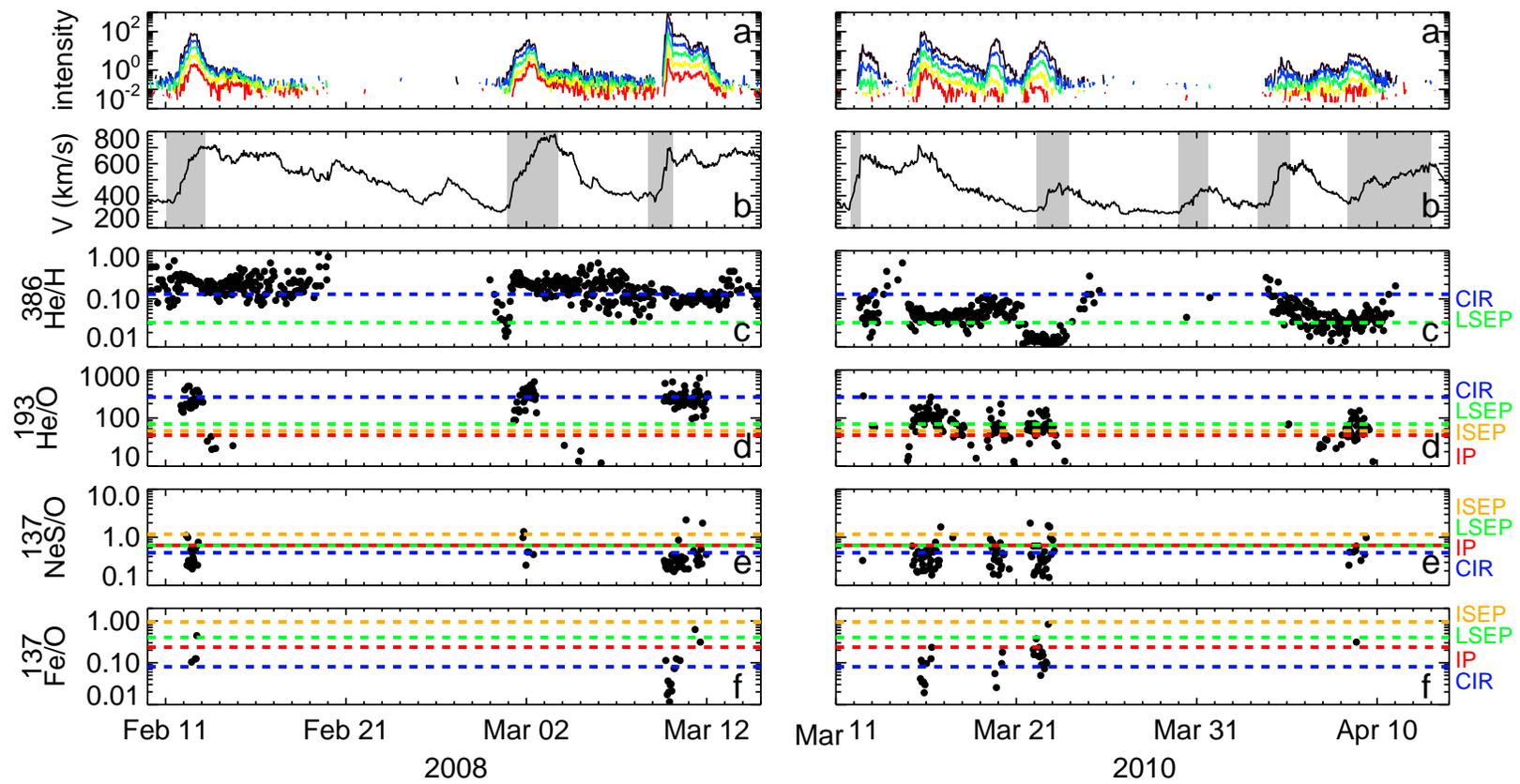